\documentstyle[12pt]{article}

\def\hybrid{\topmargin 0pt      \oddsidemargin 0pt
        \headheight 0pt \headsep 0pt

       \textwidth 6.5in        % US paper
       \textheight 9in         % US paper
        \marginparwidth 0.0in
        \parskip 5pt plus 1pt   \jot = 1.5ex}
\catcode`\@=11
\def\marginnote#1{}

\newcount\hour
\newcount\minute
\newtoks\amorpm
\hour=\time\divide\hour by60
\minute=\time{\multiply\hour by60 \global\advance\minute by-\hour}
\edef\standardtime{{\ifnum\hour<12 \global\amorpm={am}%
        \else\global\amorpm={pm}\advance\hour by-12 \fi
        \ifnum\hour=0 \hour=12 \fi
        \number\hour:\ifnum\minute<10 0\fi\number\minute\the\amorpm}}
\edef\militarytime{\number\hour:\ifnum\minute<10 0\fi\number\minute}

\def\draftlabel#1{{\@bsphack\if@filesw {\let\thepage\relax
   \xdef\@gtempa{\write\@auxout{\string
      \newlabel{#1}{{\@currentlabel}{\thepage}}}}}\@gtempa
   \if@nobreak \ifvmode\nobreak\fi\fi\fi\@esphack}
        \gdef\@eqnlabel{#1}}
\def\@eqnlabel{}
\def\@vacuum{}

\def\draftmarginnote#1{\marginpar{\raggedright\scriptsize\tt#1}}

\def\draftlabel#1{{\@bsphack\if@filesw {\let\thepage\relax
   \xdef\@gtempa{\write\@auxout{\string
      \newlabel{#1}{{\@currentlabel}{\thepage}}}}}\@gtempa
   \if@nobreak \ifvmode\nobreak\fi\fi\fi\@esphack}
        \gdef\@eqnlabel{#1}}
\def\@eqnlabel{}
\def\@vacuum{}
\def\draftmarginnote#1{\marginpar{\raggedright\scriptsize\tt#1}}

\def\draft{\oddsidemargin -.5truein
        \def\@oddfoot{\sl preliminary draft \hfil
        \rm\thepage\hfil\sl\today\quad\militarytime}
        \let\@evenfoot\@oddfoot \overfullrule 3pt
        \let\label=\draftlabel
        \let\marginnote=\draftmarginnote
   \def\@eqnnum{(\theequation)\rlap{\kern\marginparsep\tt\@eqnlabel}%
\global\let\@eqnlabel\@vacuum}  }

\def\numberbysection{\@addtoreset{equation}{section}
        \def\theequation{\thesection.\arabic{equation}}}

\def\underline#1{\relax\ifmmode\@@underline#1\else
        $\@@underline{\hbox{#1}}$\relax\fi}

\def\titlepage{\@restonecolfalse\if@twocolumn\@restonecoltrue\onecolumn
     \else \newpage \fi \thispagestyle{empty}\c@page\z@
        \def\thefootnote{\fnsymbol{footnote}} }

\def\endtitlepage{\if@restonecol\twocolumn \else  \fi
        \def\thefootnote{\arabic{footnote}}
        \setcounter{footnote}{0}}  %\c@footnote\z@ }
\relax

\numberbysection
\hybrid

\def\beq{\begin{equation}}
\def\eeq{\end{equation}}
\def\p{\partial}
\def\G{\Gamma}
\def\g{\gamma}
\def\s{\sigma}
\def\o{\omega}
\def\bea{\begin{eqnarray}}
\def\eea{\end{eqnarray}}

\newtheorem{th}{Theorem}[section]

\newtheorem{cor}{Corollary}[section]
\newtheorem{lem}{Lemma}[section]
\begin{document}

\begin{titlepage}

\title{Algebraic-geometrical n-orthogonal curvilinear coordinate systems and 
solutions to the associativity equations}

\author{I. Krichever \thanks{Department of Mathematics of Columbia
University and Landau Institute for Theoretical Physics,
Kosygina str. 2, 117940 Moscow, Russia}}

\maketitle

\begin{abstract}
Algebraic-geometrical n-orthogonal curvilinear coordinate systems in a 
flat space are constructed. They are expressed in terms of the Riemann
theta function of auxiliary algebraic curves. The exact formulae
for the potentials of algebraic geometrical Egoroff metrics and 
the partition functions of the corresponding topological field 
theories are obtained. 

\end{abstract}

\vfill

\end{titlepage}

\section{Introduction}

The problem of constructing of n-orthogonal curvilinear coordinate systems
or {\it flat} diagonal metrics
\beq
ds^2=\sum_{i=1}^n H_i^2(u) (du^i)^2, \ \ \ u=(u^1,\ldots,u^n), \label{1}
\eeq
for more than a century starting from the work of Dupin and Binet published 
in 1810 was among the first rate problems of Differential Geometry. 
As a classification problem it was mainly solved at the turn of 
this century.  A milestone in a history of this problem was a fundamental 
monograph by G. Darboux \cite{dar}.

At the beginning of the 80-s it was found that this very classical and
old problem has deep connections and applications to the modern theory
of integrable quasi-linear hydrodynamic type systems in (1+1)-dimensions
\cite{dn1}, \cite{dn2}, \cite{tzar}. 
This theory was proposed by B. Dubrovin and S. Novikov
as the Hamiltonian theory for the averaging (Whitham) equations for periodic 
solutions of integrable soliton equations in (1+1)-dimensions. 
Later it was noticed (\cite{dub}) that the classification of Egoroff 
metrics, i.e. flat diagonal metrics such that 
\beq 
\p_j H_i^2=\p_i H_j^2,\ \ \p_i={\p\over \p u^i}, \label{2} 
\eeq 
solves the classification problem of the massive topological field 
theories.  Note that (\ref{2}) implies that there exists a function $\Phi(u)$, 
called a potential, such that
\beq
H_i^2(u)=\p_i \Phi(u) \label{2a}
\eeq
 
It should be emphasized, that the "classical" results are mainly that of
the classification nature. It was shown that locally a general
solution of the Lam\'e equations  (\ref{3}), (\ref{4})
\beq
\p_k \beta_{ij}=\beta_{ik}\beta_{kj},\ \ i\neq j\neq k, \label{3} 
\eeq 
\beq 
\p_i \beta_{ij}+\p_j\beta_{ji}+\sum_{m\neq i, j} \beta_{mi}\beta_{mj}=0,
\ \ i\neq j, \label{4} 
\eeq
for, so-called, rotation coefficients
\beq
\beta_{ij}={\p_i H_j\over H_i}, \ \ i\neq j, \label{5}
\eeq
depends on $n(n-1)/2$ arbitrary functions of two variables. 
Equations (\ref{3}), (\ref{4}) are equivalent to the vanishing conditions
of all a'priory non-trivial components of the curvature tensor. 
(Equations (\ref{3}) imply that $R_{ij,ik}=0$ and (\ref{4})
imply  $R_{ij,ij}=0$.) 

If a solution for (\ref{3}), (\ref{4}) is
known, then the Lam\'e coefficients $H_i$ can be found from the linear
equations (\ref{5}) that are compatible due to (\ref{3}). They depend on 
$n$ functions of one variable that
are the initial data
\beq
f_i(u^i)=H_i(0,\ldots,0, u^i,0,\ldots,0) \label{6}
\eeq
After that flat coordinates $x^k(u)$ can be found from a set of the linear
equations
\beq
\p^2_{ij} x^k=\G_{ij}^i\p_i x^k+\G_{ji}^j \p_j x^k,  \label{7}
\eeq
\beq
\p^2_{ii} x^k=\sum_{j=1}^n \G_{ii}^j \p_j x^k, \label{8}
\eeq
where $\G_{ij}^k$ are the Christoffel coefficients of the metric (\ref{1}):
\beq
\G_{ik}^i={\p_k H_i\over H_i},\ \ 
\G_{ii}^j=-{H_i\p_i H_i\over H_j^2}, \ \ i\neq j
\label{9}
\eeq
Note that the compatibility of (\ref{7}) and (\ref{8}) requires  (\ref{3}) and  
(\ref{4}), as well.

Though this way one gets the complete description of n-th orthogonal 
systems, the list of known exact examples was relatively short. 
A number of new examples were obtained from the Whitham theory. 
In particular, in \cite{kr_t} it was shown that the moduli spaces of 
algebraic curves with jets of local coordinates
at punctures generate a wide class of flat diagonal metrics.

Recently solutions of (\ref{3}) and (\ref{4}) have been
constructed by V. Zakharov \cite{zak} with the help of the 
"dressing procedure" within the framework of the inverse problem method.
The equations (\ref{3}) are equivalent to the compatibility conditions 
for the auxiliary linear system
\beq
\p_i \Psi_j=\beta_{ij} \Psi_i, \ i\neq k \label{10}
\eeq
Therefore, any known inverse method scheme may be relatively easily adopted 
for a construction of various classes of its exact solutions. That can be the
dressing scheme or algebraic-geometrical one, as well. The next step is to 
find the way to satisfy the constraints (\ref{4}). As it was shown in \cite{zm} 
the {\it differential reduction} proposed in \cite{zak} and solving this 
problem in the case of dressing scheme becomes very natural in terms of the, 
so-called, $\bar \p$-problem. 

The main goal of this paper is not merely to construct the finite-gap
or algebraic geometrical solutions to the Lam\'e equations (\ref{3}), (\ref{4})
but to propose a scheme that solves simultaneously the whole system 
(\ref{3})-(\ref{9}), i.e. gives Lam\'e coefficients $H_i$ and flat coordinates
$x^i(u)$, as well. 

At first sight it seems that our approach is
completely different from that proposed in \cite{zak} and \cite{zm}. 
We consider the basic {\it multi-point}  Baker-Akhiezer functions $\psi(u,Q)$
which are defined by their analytical properties on auxiliary Riemann
surface $\G,\ Q\in \G$, and directly prove (without any use of differential
equations !) that under certain constraints on the corresponding set of 
algebraic-geometrical data the evaluations $x^k(u)=\psi(u,Q_k)$ of $\psi$ 
at a set of punctures on $\G$ satisfy the equations
\beq
\sum_{k,l}\eta_{kl}\p_ix^k(u) \p_j x^l(u)=0,\ \ i\neq j, \label{10a}
\eeq
where $\eta_{kl}$ is a constant matrix. 
Therefore, $x^k(u)$ are flat coordinates for the diagonal metric (\ref{1})
with the coefficients
\beq
H_i^2(u)=\sum_{k,l}\eta_{kl}\p_i x^k(u) \p_i x^l(u) \label{10b}
\eeq
It turns out that up to constant factors the Lam\'e coefficients $H_i(u)$ are 
equal to the leading terms of the expansion of the same function $\psi$ at the 
punctures $P_i$ on $\G$ where $\psi$ has exponential type singularities.
We would like to mention that our constraints on the algebraic geometrical data
that lead to solutions of (\ref{10a}), (\ref{10b}) are the generalization
of conditions proposed in (\cite{nv}) for the description of the potential
two-dimensional Schr\"odinger operators (see \cite{sit}, also).  

In the third section we link our results to the approach 
of \cite{zak}, \cite{zm} and show that $\psi$ is a {\it generating} function
\beq
\p_i \psi(u,Q)=h_i(u)\Psi_i^0(u,Q), \ \ H_i=\varepsilon_i h_i(u),\ 
\varepsilon_i=const, 
\label{10d} 
\eeq 
for solutions of the system
\bea 
\p_i \Psi^0_j & = & \beta_{ji} \Psi^0_i, \nonumber\\
\p_j \Psi^0_j & = & \Psi_j^1-\sum_{m\neq j} \beta_{mj} \Psi^0_m \label{11} 
\eea
Note that the compatibility conditions of this extended auxiliary linear system 
are equivalent to both the sets of the equations (\ref{3}) and (\ref{4}).

In the forth section of the paper we specify the algebraic geometrical
data corresponding to Egoroff metrics and obtain the exact formula
in terms of Riemann theta functions for the potentials $\Phi(u)$ of 
such metrics. 

As it was mentioned above, the connection of the classification problem
for Egoroff metrics and the classification problem of the topological field 
theories was found in \cite{dub}. The last problem for the theory
with $n$ primary fields $\phi_1,\ldots,\phi_n$ may be formulated in terms
of the {\it associativity} equations for the partition function 
$F(x_1,\ldots,x_n)$ of the deformed theories \cite{wit}, \cite {vvd}. 
These equations are the conditions that the commutative algebra with 
generators $\phi_k$ and the structure constants defined by the 
third derivatives of $F$:
\beq
c_{klm}(x)={\p^3 F(x)\over \p x^k\p x^l \p x^m}, \label{13}
\eeq
\beq \phi_k\phi_l=c_{kl}^m(x)\ \phi_m; \ c_{kl}^m=c_{kli}\ \eta^{im};\ 
\eta_{ki}\eta^{im}=\delta_{k}^{m}, \label{14} 
\eeq 
is an associative algebra, i.e.  
\beq 
c_{ij}^k(x)c_{km}^l(x)=c_{jm}^k(x)c_{ik}^l(x) \label{15} 
\eeq 
In addition, it is required that there exist constants $r^m$
such that the constant metric $\eta$ in (\ref{14}) is equal to
\beq
\eta_{kl}=r^m c_{klm}(x) \label{16}
\eeq
The conditions (\ref{15}) are a set of over-determined non-linear equations
for the function $F$. It turns out that for any solution to the system
(\ref{15}), (\ref{16}) in case when the algebra (\ref{14}) is semisimple
there exists Egoroff metric such that the third derivatives of the
partition function can be written in the form
\beq
c_{klm}=\sum_{i=1}^n H_i^2{\p u^i\over \p x^k}{\p u^i\over \p x^l} 
{\p u^i\over \p x^m}, \label{17}
\eeq
where $u^i$ and $x^k$ are the corresponding n-th orthogonal curvilinear and 
flat coordinates. Moreover, it turns out that for any set of rotation
coefficients $\beta_{ij}=\beta_{ji}$ satisfying (\ref{3}), (\ref{4})
there exists $n$-parametric family of Egoroff metrics such that
the functions defined by (\ref{17}) are the third derivatives of a certain
function $F$. (Recall, that for the given rotation coefficients there
are infinitely many corresponding metrics.)

In the last section for each algebraic geometrical Egoroff metric 
we define a function $F$ and show that its third derivatives have the form
(\ref{17}) and satisfy (\ref{15}), that are the truncated set of
the associativity conditions. We specify also the set of algebraic-geometrical 
Egoroff metrics such that (\ref{16}) is fulfilled, also, and obtain for the 
corresponding partition functions $F$ an exact formula in terms of the theta 
functions. 

\section{Bilinear relations for the Baker-Akhiezer functions
and flat diagonal metrics}

To begin with let us present some necessary facts from the 
general algebraic-geometrical scheme proposed by the author
\cite{kr1}, \cite{kr2}. The core-stone of this scheme is a notion of the 
Baker-Akhiezer functions that are defined by their analytical 
properties on the auxiliary Riemann surfaces.

Let $\G$ be a smooth genus $g$ algebraic curve with fixed local coordinates
$w_i(Q)$ in neighborhoods of $n$ punctures $P_i,\ i=1,\ldots,n,$ on $\G$,
$w_i(P_i)=0$. Then for any set of $l$ points $R_{\alpha},\ \alpha=
1,\ldots,l,$ and for any set of $g+l-1$ 
points $\g_1,\ldots,\g_{g+l-1}$ in a general position there exists a unique 
function $\psi(u,Q|D,R), \ u=(u_1,\ldots,u_n), Q\in \G$, such that:

$1^0.$ $\psi(u,Q|D,R)$ as a function of the variable $Q\in \G$ is
meromorphic outside the punctures $P_j$ and at most has simple poles at 
the points $\g_s$ (if all of them are distinct);

$2^0.$ in the neighborhood of the puncture $P_j$ the function $\psi$ 
has the form:
\beq
\psi=e^{u^j w_j^{-1}}
(\sum_{s=0}^{\infty}\xi_{s}^{j}(u)w_j^s),\ w_j=w_j(Q);
\label{1.1}
\eeq

$3^0.$ $\psi$ satisfies the conditions
\beq 
\psi(u,R_{\alpha})=1 \label{1.0} \eeq

\noindent 
Below we shall often denote the Baker-Akhiezer function by $\psi(u,Q)$
without explicit indication on the defining divisors $D=\g_1+\cdots+\g_{g+l-1}$ 
and $R=R_1+\cdots+R_l$.   

The exact theta-functional formula for the Baker-Akhiezer functions in
terms of the Riemann theta-functions was proposed (\cite{kr2}) as a 
generalization of the formula found in \cite{mi} for the Bloch solutions of the 
finite-gap ordinary Schr\"odinger operators. 

The Riemann theta-function corresponding to an algebraic genus $g$ curve $\G$
is an entire function of $g$ complex variables $z=(z_1,\ldots,z_g)$ 
defined by the formula
\beq
\theta(z_1,\ldots,z_g)=\sum_{m\in Z^g}e^{2\pi i(m,z)+\pi
i(Bm,m)}, \label{1.4} 
\eeq
where the matrix $B=B_{ij}$ is a matrix of $b$-periods  
\beq
B_{ij}=\oint_{b_i}\omega_j \label{1.5}
\eeq
of the normalized holomorphic differentials $\omega_j(P)$ 
\beq 
\oint_{a_j} \omega_i=\delta_{ij} \label{1.6}
\eeq
on $\G$. Here $a_i,b_i$ is a basis of cycles on $\G$ with canonical 
matrix of intersections $a_i\cdot a_j=b_i\cdot b_j=0, \ a_i\cdot
b_j=\delta_{ij}$.

The theta function has remarkable automorphy properties with respect
to the lattice ${\cal B}$ generated
by the basic vectors $e_i\in C^g$ and by the vectors $B_j\in C^g$ with
coordinates $B_{ij}$: for any $l\in{\bf Z}^g$ and $z\in{\bf C}^g$
\beq
\theta(z+l)=\theta(z),\ \ 
\theta(z+Bl)= \exp [-i\pi (Bl,l)-2i\pi (l,z)]\theta(z)\label{1.7}
\eeq
The torus  $J(\Gamma)$  
\begin{equation}
J(\Gamma)=C^g/{\cal B} 
\end{equation}
is called the Jacobian variety of the algebraic curve $\G$.

The vector $A(P)$ with coordinates
\beq
A_k(Q)=\int_{q_0}^Q \omega_k \label{1.8}
\eeq
defines the, the so-called,  Abel map. 

According to the Riemann-Roch theorem for any divisors 
$D=\gamma_1+\cdots+\gamma_{g+l-1}$ and $R=R_1+\cdots+R_l$ in the
general position there exists a unique meromorphic function 
$r_{\alpha}(Q)$ such that the divisor of its poles coincides 
with $D$ and such that 
\beq
r_{\alpha}(R_{\beta})=\delta_{\alpha, \beta} \label{1.9} 
\eeq
This function may be written as follows (see \cite{bab}):  
\beq
r_{\alpha}(Q)={f_{\alpha}(Q)\over f_{\alpha}(R_{\alpha})}; \quad 
f_{\alpha}(Q)=\theta(A(Q)+Z_{\alpha}) {\prod_{\beta\neq \alpha}
\theta(A(Q)+F_{\beta})\over \prod_{m=1}^l\theta (A(Q)+S_m)}, \label{1.10} 
\eeq 
where 
\beq F_{\beta}=-{\cal K}-A(R_{\beta})-\sum_{s=1}^{g-1} A(\gamma_s), 
\label{1.10a}
\eeq 
\beq 
S_m=-{\cal K}-A(\gamma_{g-1+m})-\sum_{s=1}^{g-1} A(\gamma_s), \label{1.10b}
\eeq 
\beq 
Z_{\alpha}=Z_0-A(R_{\alpha}), \quad Z_0=-{\cal K}-\sum_{s=1}^{g+l-1} 
A(\gamma_s)+\sum_{\alpha=1}^l A(R_{\alpha}). \label{1.11} 
\eeq
and ${\cal K}$ is a vector of the Riemann constants.

Let $d\Omega_j$ be the unique meromorphic 
differential holomorphic on $\G$ outside the puncture $P_j$, 
which has the form 
\beq 
d\Omega_j=d(w_j^{-1}+O(w_j)) \label{1.12} 
\eeq 
near the puncture and is normalized by the conditions 
\beq 
\oint_{a_k}d\Omega_j=0 \label{1.13} 
\eeq 
It defines a vector $V^{(j)}$ 
with coordinates 
\beq 
V^{(j)}_k={1\over 2\pi i} \oint_{b_k} d\Omega_j \label{1.14} 
\eeq 
\begin{th} The Baker-Akhiezer function $\psi(u,Q|D,R))$ has the form:
\beq
\psi=\sum_{\alpha=1}^l
r_{\alpha}(Q){\theta (A(Q)+\sum_{i=1}^n (u^i V^{(i)})+Z_{\alpha}) 
\theta(Z_0)\over \theta (A(Q)+Z_{\alpha})) \theta (\sum_{i=1}^n (u^i 
V^{(i)})+Z_0) } \exp{\left(\sum_{i=1}^n 
u^i\int_{R_{\alpha}}^Qd\Omega_i\right)} \label{1.15} \eeq 
\end{th}

\bigskip
\noindent{\bf Admissible curves.} 

\medskip
\noindent Now in order to get the algebraic-geometrical flat diagonal metrics 
and their flat coordinates we are going to specify the algebraic-geometrical 
data defining the Baker-Akhiezer functions. The corresponding algebraic curve 
$\G$ should be a curve with a holomorphic involution 
\beq
\sigma:\G \to \G, \label{2.1}
\eeq
with $2m\geq n$ fixed points $P_1,\ldots,P_{n},Q_1,\ldots,Q_{2m-n}$, $m\leq n$. 
The local coordinates $w_j(Q)$ in the neighborhoods of $P_1,\ldots,P_n$ 
should be odd 
\beq
w_j(Q)=-w_j(\sigma(Q)).\label{2.2}
\eeq 
The factor curve $\G_0=\G/ \sigma$ is a smooth algebraic curve.
The projection
\beq
\pi :\Gamma \longmapsto \Gamma_0=\Gamma / \sigma \label{2.3a}
\eeq
represents $\Gamma$ as a two-sheet covering of $\Gamma_0$ with $2m$
branching points $P_j,\ Q_s$. In this realization the involution $\sigma$ is
a permutation of the sheets. For $Q\in \G$ we denote the point
$\sigma(Q)$ by $Q^{\sigma}$.

From the Riemann-Hurwitz formula it follows that
\beq
g=2g_0-1+m, \label{2.3}
\eeq
where $g_0$ is genus of $\Gamma_0$.

In other words to construct an admissible algebraic curve $\G$
one has to start with an algebraic curve $\G_0$, take a meromorphic function
$E(P), \ P\in \G_0,$ with $2m$ simple zeros or poles on it
and take $\G$ as a Riemann surface of the function $\sqrt{E(P)}$. 
If $w_j^0 (P)$ are local coordinates on $\G_0$ at the points $P_j$ then we 
choose $w_j=\sqrt{w_j^0(P)}$ as the local coordinates on $\G$ at the branching 
points $P_j$.

\bigskip

\noindent{\bf Admissible divisors.} 

\medskip
\noindent Let us fix on $\G_0$ a set of $n-m$ punctures 
$\hat Q_1,\ldots,\hat Q_{n-m}$. 
A pair of the divisors $D$ and $R$ on $\G$ is called admissible if
there exists a meromorphic differential $d\Omega_0$ on $\G_0$ such that:

a) $d\Omega_0(P), \ P\in \G_0$ has  $m+l$ simple poles at the points 
$Q_1,\ldots, Q_{2m-n}$, $\hat Q_1,\ldots,\hat Q_{n-m}$ and at the 
points $\hat R_{\alpha}=\pi(R_{\alpha})$; 

b) the differential $d\Omega_0$ is equal to zero at the projections 
$\hat \g_s$ of the points of the divisor $D$,
\beq 
d\Omega_0(\hat \gamma_s)=0, \ \ \hat \g_s=\pi(\g_s) \label{2.5} 
\eeq 

The differential $d\Omega_0$ can be considered an even (with respect to
involution $\sigma$) meromorphic differential on $\G$ where it has
$n+2l$ simple poles at the branching points $Q_1,\ldots,Q_{2m-n}$ and at
the preimages of its other poles on $\G_0$. Let us denote the preimages of
the points $\hat Q_k$ by
$Q_{2m-n+1},\ldots, Q_{2m}$:
\beq
\pi(Q_{2m-n+i})=\pi(Q_{n-i+1})=\hat Q_i,\ i=1,\ldots,n-m. \label{2.7}
\eeq
The involution $\sigma$ induces the involution $\sigma(k)$ of the indices
of the punctures $Q_k,\\ \sigma(Q_k)=Q_{\sigma(k)})$:
\beq
\sigma(k)=k,\ k=1,\ldots, 2m-n;\ \ \sigma(k)=2m-k+1,\ k=2m-n+1,\ldots, n 
\label{2.70}
\eeq
In terms of equivalence classes the admissible pairs of the divisors 
$D$ and $R$ can be described as that satisfying the condition:  
\beq
D+D^{\sigma}-R-R^{\s}=K+\sum_{j=1}^n (Q_j-P_j). \label{2.7a} \eeq

\medskip

\noindent{\it Example. Hyperelliptic curves.}

\medskip
\noindent The simplest example of the admissible curve is the hyperelliptic
curve $\G$ defined by the equation 
\beq 
\lambda^2={\prod_{j=1}^{2m-n} (E-Q_j) \prod_{k=1}^{n-m}(E-\hat Q_k)^2
\over \prod_{i=1}^n(E-P_i)}, \  m\leq n \label{3.1} 
\eeq 
Here $P_i,\ Q_j, \ \hat Q_k$ are complex numbers. A point $\g$ on $\G$ 
is a complex number and a sign of the square root from the right hand side of 
(\ref{3.1}).  The curve $\G$ has genus $g=m-1$. Any set of $m+l-2$ points 
$\g_s,\ \g_s\neq \g_{s'}$ and any set of $l$ points are a admissible pair of
the divisors. The corresponding differential $d\Omega_0$ is equal to 
\beq
d\Omega_0={\prod_{s=1}^{m+l-2} (E-\g_s)\over \prod_{j=1}^{2m-n} (E-Q_j) 
\prod_{k=1}^{n-m}(E-\hat Q_k)\prod_{\alpha=1}^l (E-R_{\alpha})}dE 
\label{3.2} 
\eeq 
As we shall see in the last section of the paper the flat diagonal metrics 
corresponding to the hyperelliptic curves are Egoroff metrics. Moreover,
this set of data gives also the simplest examples of algebraic-geometrical
solutions to the associativity equations. 

\medskip

\noindent{\bf Important remark.} 
In the following main body of the paper we assume for the simplicity of the 
formulae, only, that the divisors $R$ and the punctures $Q_j$ are in the 
general position and do not intersect with each other.  We return to the special 
case when they coincide at the end of the last section.

\begin{th} Let $\psi(u,Q|D,R)$ be the Baker-Akhiezer function
corresponding to an admissible algebraic curve and 
an admissible pair of the divisors $D$ and $R$. Then the functions $x^j(u)$:
\beq         
x^j(u)=\psi(u,Q_j),\ \ j=1,\ldots, n, \label{2.8}
\eeq
satisfy the equations
\beq
\sum_{k,l}\eta_{kl}\p_i x^k \p_j x^l = \varepsilon_i^2 h_i^2 \delta_{ij}, 
\label{2.9} 
\eeq 
where 
\beq
h_i=\xi^i_0(u) \label{2.8a}
\eeq 
are the first coefficients of the expansions (\ref{1.1}) of $\psi$ at the 
punctures $P_i$; the constants $\varepsilon_i^2$ are defined by the expansion  
of $d\Omega_0$ at the punctures $P_i$  
\beq 
d\Omega_0={1\over 2}(\varepsilon_i^2+O(w_i^0))dw_i^0=
w_i(\varepsilon_i^2+O(w_i^2))dw_i \label{2.11} 
\eeq
and at last the constants $\eta_{kl}$  are equal to 
\beq 
\eta_{kl} = \eta_k \delta_{k,\sigma(l)},\ 
\eta_k={\rm res}_{Q_k}d\Omega_0,  
\label{2.10} 
\eeq
where $\sigma(k)$ is the involution of indices defined in (\ref{2.70}).
\end{th}
{\it Proof.} Let us consider the differential 
\beq 
d\Omega^{(1)}_{ij}(u,Q)=\p_i\psi(u,Q)\ \p_j\psi(u,\sigma(Q)) 
d\Omega_0(\pi (Q)) 
\label{2.12} 
\eeq
From the definition of the admissible data, it follows that this differential
for $i\neq j$ is a meromorphic differential with poles at the points 
$Q_1,\ldots,Q_n$, only. Indeed, the poles of the first two factors 
$\p_i\psi_i(u,Q)$ and $\p_j\psi(u,\sigma(Q)$ at the points $\g_s$ and 
$\sigma(\g_s)$ cancel with zeros of $d\Omega_0$. 
At the punctures $P_k$ the essential singularities of 
the same factors cancel each other. The product of these factors has simple
poles at the points $P_i$ and $P_j$ that cancel with zeros of $d\Omega_0$
considered as a differential on $\G$, (\ref{2.11}). At last $d\Omega^{(1)}$ has
no poles at the points $R_{\alpha}$ and $R_{\alpha}^{\sigma}$  due to the 
condition (\ref{1.0}). A sum of all the residues of a meromorphic differential 
on a compact Riemann surface is equal to zero. Therefore,
\beq
\sum_{k=1}^n {\rm res}_{Q_k} d\Omega^{(1)}_{ij} =0 , \ \ i\neq j \label{2.13}
\eeq
The left hand side of this equality coincides with the left hand side
of (\ref{2.9}). 

In the case $i=j$, the differential $d\Omega_{ii}^{(1)}$ has an additional pole
at the point $P_i$ with the residue
\beq
{\rm res}_{P_i} d\Omega^{(1)}_{ii}=-\varepsilon_i^2 h_i^2 
\label{2.14} 
\eeq 
That implies (\ref{2.9}) for $i=j$ and completes the proof of the 
theorem.
\begin{cor} Let $\{\G,P_i,Q_j, D, R\}$ be a set of the admissible data. 
Then the formula 
\beq 
H_i(u)=\varepsilon_i \ \sum_{\alpha=1}^l r_{\alpha}(P_i)
{\theta (A(P_i)+\sum_{i=1}^n (u^i V^{(i)})+Z_{\alpha}) \theta(Z_0)
\over \theta (A(P_i)+Z_{\alpha})
\theta (\sum_{i=1}^n (u^i V^{(i)})+Z_0) }\  
\exp{\left(\sum_{j=1}^n \omega_{ij}^{\alpha} u^j\right)}, 
\label{2.15} 
\eeq
where $r_{\alpha}(Q)$ is the function defined by (\ref{1.10}),
\bea
\omega_{ij}^{\alpha}&=&\int_{R_{\alpha}}^{P_i} d\Omega_j,\ i\neq j,\\ 
\omega_{ii}^{\alpha}&=&\lim_{Q\to P_i}\left(\int_{R_{\alpha}}^Q d\Omega_i -
w_i^{-1}(Q)\right), \label{2.16}
\eea
defines the coefficients of the flat diagonal metric. The  
flat coordinates (not orthonormal) corresponding to this metric are given by 
the formulae:  
\beq 
x^k(u)=\sum_{\alpha=1}^lr_{\alpha}(Q_k)\ x_{\alpha}^k(u), \label{2.17a}
\eeq 
\beq
x_{\alpha}^k(u)=
{\theta (A(Q_k)+\sum_{i=1}^n (u^i V^{(i)})+Z_{\alpha})\theta(Z_0) 
\over \theta (A(Q_k)+Z_{\alpha}) 
\theta (\sum_{i=1}^n (u^i V^{(i)})+Z_0) }
\exp{\left(\sum_{i=1}^n u^i\int_{R_{\alpha}}^{Q_k}d\Omega_i\right)}
\label{2.17} 
\eeq
\end{cor}
{\bf Important remark.} Note that the theta-function and the vectors $V^{(j)}$ 
are defined by the curve $\G_0$, the punctures $\{P_1,\ldots,P_n,\ 
Q_1,\ldots,Q_{2m-n}\}$ and the first jets of local coordinates at $P_j$. A 
space of these parameters for given genus $g_0$ of $\G_0$ is equal to 
$3g_0-3+2m+n$. The vectors $Z_{\alpha}$ are defined in (\ref{1.11}), where the 
vector $Z_0$ depends on the points $\hat Q_k,\ k=1,\ldots,n-m$ and the
divisors $D$, $R$ also. It belongs to the affine space defined by the relation 
(\ref{2.7a}). We would like to stress that in the formula (\ref{2.15}) the last 
set of parameters enters through $\varepsilon_i$ and $Z_0$, only. Therefore, for 
$(n-m)>g_0$ formula (\ref{2.15}) with an {\it arbitrary} vector $Z_0$ and 
with {\it arbitrary} constants $\varepsilon_i$ defines the flat diagonal 
metric.

\bigskip
\noindent{\bf Reality conditions.} 

\medskip
\noindent
In a general case the above-constructed flat diagonal metrics $H_i(u)$ and 
their flat coordinates are complex meromorphic functions of the variables 
$u^i$. Let us specify the algebraic geometrical data such that the 
coefficients of the corresponding metrics are {\it real} functions of the {\it 
real} variables $u^i$.

Let $\G_0$ be a real algebraic curve, i.e. a curve with an antiholomorphic
involution
\beq
\tau_0:\ \G_0\longmapsto \G_0 \label{2.18}
\eeq
and let the punctures $\{P_1,\ldots,P_n\}$ and $\{Q_1,\ldots,Q_{2m-n}\}$ be
fixed points of $\tau_0$. Then there exists an antiholomorphic
involution $\tau$ on $\G$. We assume that the local coordinates $w_j$ at
the punctures $P_j$ satisfy the condition
\beq
w_j(\tau(Q))=\overline {w_j(Q)}   \label{2.19}
\eeq
Let us assume that the set of the points $\hat Q_k$ and the 
divisors $D$, $R$ are invariant with respect to $\tau$, also:
\beq 
\tau(Q_j)=Q_{\kappa(j)},\ \tau(R_{\alpha})=R_{\kappa_1(\alpha)},\ 
\tau(\g_s)=\g_{\kappa_2(s)}, \label{2.22} 
\eeq
where $\kappa_i(.)$ are the corresponding permutations of indices. 
The conditions (\ref{2.22}) imply that
the differentials $d\Omega_0$ corresponding to the
admissible pair $D$, $R$ satisfies the condition 
\beq \tau_0^*\ 
d\Omega_0=\overline{d\Omega_0}\label{2.20} \eeq 
\begin{th} Let a set of the admissible data be real (i.e. satisfies the
conditions (\ref{2.22})). Then the Baker-Akhiezer function $\psi(u,Q|D,R)$ 
satisfies the relation 
\beq
\psi (u,Q|D,R) = \overline{\psi(u,\tau(Q)|D,R)} \label{2.23} 
\eeq 
and the formula (\ref{2.15}) defines a real flat diagonal metric.  
\end{th} 
The signature of the corresponding metrics depends on the involutions 
$\kappa(j)$ and $\kappa_1(\alpha)$. For various  choices of the initial data 
one may get the flat diagonal metrics in all the pseudo-euclidain spaces 
$R^{p,q}$. In general, these metrics  are singular at some values of the 
variables $u^i$. In order to get smooth metrics for all the values of $u$ it is 
necessary to further specify the initial data. The way to do that is 
relatively standard in the finite-gap theory and we address this problem 
somewhere else.

\section{Differential equations for the Baker-Akhiezer function}

In this section we are going to clarify the meaning of our constraints
in terms of differential equations for the Baker-Akhiezer functions.
First of all, we discuss the equations for an arbitrary set of
algebraic geometrical data.

The following statement is a simple generalization of the results \cite{ndk}
where in the case $n=2$ it was shown that the corresponding Baker-Akhiezer
function satisfies the two-dimensional Schr\'odinger type equation.
\begin{lem} Let $\psi(u,Q|D,R)$ be the Baker-Akhiezer 
function. Then it satisfies the equation:
\beq
\p_i\p_j \psi = c_{ij}^i\p_i \psi+c_{ij}^j \p_j \psi, \ \ i\neq j, \label{1.2}
\eeq
where
\beq
c_{ij}^i(u)={\p_j h_i\over h_i},\ \ c_{ij}^j(u)={\p_i h_j\over h_j} 
\label{1.2a}
\eeq
and 
\beq
h_i(u)=\xi_0^i(u), \label{1.2b}
\eeq
are the first coefficients of the expansion (\ref{1.1}).
\end{lem}
The equations (\ref{1.2}) have the form (\ref{7}) 
that are the first set of the equations defining flat coordinates for 
the diagonal metric with the coefficients $H_i(u)=\varepsilon_ih_i(u)$,
where $\varepsilon_i$ are constants. Now we are going to obtain the 
equations that in the case of admissible algebraic-geometrical data 
provide the second set (\ref{8}) of the equations for the flat coordinates.

Let $\{\G, \ P_j, \  w_j , \g_s,\ R_{\alpha} \}$ be a set of the
data that defines the Baker-Akhiezer functions $\psi(u,Q|D,R)$. Let us fix 
a set of additional $n$ points $Q_1,\ldots,Q_n$. Then their exists a unique 
function $\psi^1=\psi^1(u,Q|D,R)$ such that:

$1^1.$ $\psi^1(u,Q)$, as a function of the variable $Q\in \G$, is
meromorphic outside the punctures $P_j$, at most has simple poles at 
the points $\g_s$ and equals to zero at the punctures $Q_1,\ldots,Q_n$:
\beq
\psi^1(u,Q_k)=0; \label{2.180}
\eeq

$2^1.$ in the neighborhood of the puncture $P_j$ the function $\psi^1$ 
has the form:
\beq
\psi^1=w_j^{-1}e^{u^j w_j^{-1}}
(\sum_{s=0}^{\infty}\xi_{1,s}^{j}(u)w_j^s),\ w_j=w_j(Q); 
\label{2.181}
\eeq

$3^1.$ at the points $R_{\alpha}$ the function $\psi^1$ equals $1$,
\beq
\psi^1(u,R_{\alpha}|D,R)=1 \label{2.190}
\eeq
\begin{lem} Let $\psi(u,Q|D,R)$ and $\psi^1(u,Q|D,R)$
be the Baker-Akhiezer functions corresponding to an arbitrary set of the 
algebraic-geometric data. Then they satisfy the equations 
\beq 
\p_i^2 \psi -c_i^1 \p_i \psi^1+\sum_{j=1}^nv_{ij}\p_j\psi=0, 
\label{2.200} 
\eeq 
where 
\beq 
c_i^1={h_i\over h_i^1} ,\ \ 
v_{ii}={\p_i h_i^1\over h_i^1}-2{\p_i h_i\over h_i}+
{g_i^{1}\over h_i^1}- {g_i\over h_i} \label{2.210} 
\eeq 
\beq 
v_{ij}={h_i\over h_j}\ {\p_i h_j^1\over h_i^1},\ i\neq j,\label{2.220} 
\eeq 
and the functions 
\beq h_i=\xi_0^i,\ \ h_i^1=\xi_{1,0}^i, \ g_i=\xi_1^i,\ 
g_i^1=\xi_{1,1}^i \label{2.230} 
\eeq are the first coefficients of the expansions (\ref{1.1}) and 
(\ref{2.181}).  
\end{lem} 
Again the proof is standard. Consider the function 
that is defined by the left hand side of (\ref{2.200}). The formulae 
(\ref{2.210}) and (\ref{2.220}) imply that this function satisfies all the 
defining conditions of the function $\psi$ and is regular at all the punctures 
$R_{\alpha}$. Therefore, it is equal to zero.

Now consider the case of the admissible algebraic-geometrical data. (In that 
case the set of the punctures in the definition of $\psi^1$ is the same
set as in the definition of the admissible curves and divisors, i.e.
$Q_1,\ldots,Q_{2m-n}$ are the branching points and $Q_{2m-n+1},\ldots, Q_{2n}$
are the preimages of $\hat Q_k$.)

\begin{th} Let $\psi(u,Q|D,R)$ and $\psi^1(u,Q|D,R)$ be the 
Baker-Akhiezer functions corresponding to an admissible set of 
algebraic-geometrical data. Then the equations (\ref{1.2}) and the equations
\beq 
\p_i^2 \psi =c_i^1 \p_i \psi^1+\sum_{j=1}^{n}\G_{ii}^j\p_j\psi \label{2.44}
\eeq
are fulfilled. Here $\G_{ii}^j$ are the Cristoffel's coefficients (\ref{9})
of the metric $H_i(u)=\varepsilon_i h_i(u)$.
\end{th}
{\it Proof.} 
Let us consider the differential
\beq
d\Omega^{(2)}_{ij}=\p_i\psi^1(u,Q)\ \p_j\psi(u,Q^{\sigma}) 
d\Omega_0(\pi (Q)) 
\label{2.46} 
\eeq
This differential is holomorphic everywhere except for the points $P_i$ and 
$P_j$.
The residues at these points are equal to,
\beq
{\rm res}_{P_i} d\Omega_{ij}^{(2)}=\varepsilon_i^2 h_i^1\p_j h_i,\ \ 
{\rm res}_{P_j} d\Omega_{ij}^{(2)}=-\varepsilon_j^2 h_j\p_i h_j^1.
\label{2.47}
\eeq
Therefore,
\beq
\varepsilon_i^2 h_i^1\p_j h_i=\varepsilon_j^2 h_j\p_i h_j^1. \label{2.48}
\eeq
The last formula implies that the coefficients $v_{ij}$ (\ref{2.220})
are equal to $\G_{ii}^j$ for $i\neq j$.

The differential $d\Omega_{ii}^1$ has the only pole at the point $P_i$.
Therefore, its residue at this point is equal to zero,
\beq
{\rm res}_{P_i}d\Omega_{ii}^1=h_i^1(g_i+\p_ih_i)-h_i(g_i^1+\p_ih_i^1)=0.
\label{2.49}
\eeq
(\ref{2.49}) implies that $v_{ii}$ given by (\ref{2.210}) are equal to
$\G_{ii}^i$. 

Note once again, that at the points $Q_j$ (where $\psi^1$ is equal to zero)
the equations (\ref{2.44}) coincide with (\ref{8}).
\begin{cor} The functions $\Psi_i^0(u,Q)$ and $\Psi_i^1(u,Q)$ 
\beq 
\Psi_i^0(u,Q)={1\over h_i(u)}\p_i \psi(u,Q),\ \ 
\Psi^1_i(u,Q)={1\over h_i^1(u)}\p_i \psi^1(u,Q). \label{2.50}
\eeq
satisfy the equations (\ref{11}), where $\beta_{ij}(u)$ are the rotation
coefficients (\ref{5}) of the metric $H_i(u)$.
\end{cor}
The proof of the corollary follows from the direct substitution of
(\ref{2.50}) into (\ref{1.2}) and (\ref{2.44}).

Consider the analytical properties of $\Psi_i^0(u,Q)$ and 
$\Psi_i^1(u,Q)$ as functions on the algebraic curve $\G$. The defining
analytical properties of the Baker-Akhiezer function imply that:

$1^{*}.$ $\Psi_i^N(u,Q),\ N=0,1$ is meromorphic outside the punctures 
$P_j$ and at most has simple poles at the points $\g_1,\ldots,\g_{g+l-1}$;

$2^{*}.$ in the neighborhood of the puncture $P_j$ the function $\Psi_i^N$ 
has the form:
\beq
\Psi_i^N=w_j^{-N-1}e^{u_j w_j^{-1}}
(\delta_{ij}+\sum_{s=1}^{\infty}\zeta_{s,N}^{ij}(u)w_j^s),\ w_j=w_j(Q); 
\label{2.60}
\eeq

$3^{*}.$ the functions $\Psi_i^N$ are equal to zero at the punctures 
$R_{\alpha}$:
\beq
\Psi_i^N(u,R_{\alpha})=0; \label{2.60a}  
\eeq

$4^{*}.$ the functions $\Psi_i^1$ are equal to zero at the punctures $Q_j$:
\beq 
\Psi_i^1(u,Q_j)=0 \label{2.61} 
\eeq 
\begin{lem} Let $\G$ be a smooth genus $g$ algebraic curve with 
$2n$ punctures $P_j,\ Q_j$ and with fixed local coordinates 
$w_j(Q)$ in the neighborhoods of the punctures $P_j$. Then for any set of 
$(g+l-1)$ point $\g_s$ in a general position there exist the unique functions 
$\Psi_i^0(u,Q)$ and $\Psi_i^1(u,Q)$ satisfying the above formulated conditions 
$1^{*}$-$3^{*}$.  
\end{lem} 
For a given admissible curve $\G$ with punctures $P_i, \ Q_i$ and local 
coordinates $w_i$ the Baker-Akhiezer functions and therefore, the coefficients
$H_i(u|D,R)$ of the corresponding diagonal flat metric depend on the admissible
pair of the divisors $D$ and $R$. Two pairs of the divisors $D,R$ and $D',R'$ 
are called equivalent if there exists a meromorphic function $f(Q)$ on $\G$ 
such the divisor $(f)_{\infty}$ of its pooles and the divisor $(f)_0$ of its 
zeros are equal to 
\beq (f)_{\infty}=D+R' ,\ \ (f)_0=D'+R \label{2.600} \eeq 
In terms of the Abelian map $D,R$ are equivalent to $D',R'$ iff 
\beq 
\sum_{s=1}^{g+l-1}A(\g_s)-\sum_{\alpha=1}^l A(R_{\alpha})=
\sum_{s=1}^{g+l'-1}A(\g_s)-\sum_{\alpha=1}^{l'} A(R_{\alpha}')\label{2.61a} 
\eeq
Lemma 3.3 implies that the following statement is valid.  \begin{cor}
The rotation coefficients $\beta_{ij} (u|D,R)$ and $\beta_{ij}(u,|D'R')$
corresponding to the equivalent pairs of the divisors satisfy the relation
\beq
f(P_i)\beta_{ij}(u|D,R)=f(P_j) \beta_{ij}(u|D'R'), \label{2.601}
\eeq
where $f(Q)$ is the function such that (\ref{2.600}) are fulfilled.
\end{cor}
From the uniqueness of $\Psi_i^N$ and the definition of the equivalent
pairs it follows that
\beq
f(P_i)\Psi_i^N(u,Q|D,R)=f(Q)\Psi_i^N(u,Q|D',R') \label{2.602}
\eeq
and, therefore, (\ref{2.601}) are valid.

Let us express the rotation coefficients and the Lam\'e coefficients in terms
of the function $\Psi_i^1(u,Q|D,R)$, only.
\begin{th} 
The rotation coefficients $\beta_{ij}(u)$ of the algebraic-geometrical flat 
diagonal metric with the coefficients $H_i(u|D,R)$ corresponding to the
Baker-Akhiezer function \\
$\psi(u,Q|D,R)$ are equal to 
\beq 
\beta_{ij}=\beta_{ij}(u|D,R)=\zeta_{1,1}^{ji}(u|D,R), \label{2.63} \eeq 
where $\zeta_{1,1}^{ji}$ are the first coefficients of the expansion 
(\ref{2.60}) for the functions $\Psi_i^1(u,Q|D,R)$. 
The Lam\'e coefficients $H_i(u|D,R,r')$ are equal to 
\beq
H_i(u|D,R)=-\sum_{\alpha} d_{\alpha} \Psi_i^1 (u,R_{\alpha}^{\sigma}|D,R), 
\label{2.64} 
\eeq 
where
\beq 
d_{\alpha}={\rm res}_{R_\alpha} d\Omega_0     \label{2.640}
\eeq
\end{th} 
{\it Proof.} The equations (\ref{11}) imply that the functions $\Psi_i^1$ 
satisfy the equation
\beq
\p_i \Psi_j^1=\beta_{ij} \Psi^1_i, \ \ i\neq j \label{2.605}
\eeq
The formula (\ref{2.63}) is a result of the direct substitution of the 
expansion (\ref{2.60}) into (\ref{2.605}). For the proof of (\ref{2.64}) 
let us consider the differential 
\beq 
d\Omega_{i}^{(3)}(u,Q)=\Psi_i^1(u,Q)\psi(u,Q^{\sigma}) d\Omega_0 \label{2.65} 
\eeq 
This differential is meromorphic with the poles at the point $P_i$ and the 
punctures $R_{\alpha}^{\s}$. The residue of $d\Omega_i^{(3)}$ at the point $P_i$ 
is equal to \beq {\rm res}_{P_i}d\Omega_i^{(3)}=H_i(u). \label{2.66} \eeq The 
residues of this differential at the points $R_{\alpha}^{\sigma}$ are equal to 
the corresponding terms of the sum in the right hand side of (\ref{2.64}). The 
sum of all these residues is equal to zero. 
Theorem is proved.

\noindent{\bf Important remark.} Theorem 3.2 allows one to give another
interpretation of our basic construction. The rotation
coefficients $\beta_{ij}$ up to gauge transformation (\ref{2.601}), depend
on the equivalence classes of the divisors $D,R$, only. The 
functions $\Psi_i^1(u,Q|D,R)$ satisfy (\ref{2.605}). Therefore, any linear 
combination of the form (\ref{2.64}) defines the coefficients of the flat 
diagonal metrics. Therefore, the formulae (\ref{2.17}) may be considered as the
solution for the problem of the construction of the corresponding flat
coordinates.

\section{Egoroff metrics and the partition functions of the
topological field theories}

\noindent In this section we describe the algebraic geometrical data 
corresponding to Egoroff metric, i.e. the metrics with symmetric rotation 
coefficients $\beta_{ij}=\beta_{ji}$. 

Let $E(P)$ be a meromorphic function on a smooth genus $g_0$ algebraic curve
$G_0$ with $n$ simple poles at points $P_i$ having $2m-n$ simple zeros
at points $Q_1,\ldots, Q_{2m-n}$ and $n-m$ double zeros at points
$\hat Q_1,\ldots, \hat Q_{n-m}$. The Riemann surface $\G$ of the
function
\beq
\lambda=\sqrt{E(P)} \label{5.1}
\eeq
is an admissible curve in the sense of our definition. The point $Q\in G$
is a pair $P$ and a sign $\pm$ of the square root. The function
$\lambda=\lambda(Q)$ is an odd function with respect to the involution of $\G$. 
As a function on $\G$ it has simple poles at the points $P_i$ and simple zeros 
at the points $Q_j,\ j=1,\ldots,n$. The function $\lambda^{-1}$ defines local 
coordinates 
\beq
w_j(Q)=\lambda^{-1}(Q) \label{5.2} 
\eeq
in the neighborhoods of the punctures $P_i$.  

\medskip
\noindent{\it Example.} Let $\G_0$ be a plane curve defined by the
equation
\beq
L(y,E)=0, \label{5.2a}
\eeq
where $L$ is degree $n$ polynomial in two variables. Let
$\G_0$ has $n$ smooth points $P_j$ and $n$ smooth points
$Q_j$ as preimages of $E=\infty$ and $E=0$, respectively. Then the admissible 
curve $\G$ is defined by the equation
\beq
L(y,\lambda^2)=0 \label{5.2b}
\eeq
\begin{th}
Let $D,R$ be an admissible pair of the divisors on the Riemann surface $\G$ of 
the function $\lambda(Q)$. Then the rotation coefficients defined by $D,R$ are 
symmetric 
\beq 
\beta_{ij}(u|D,R)=\beta_{ji}(u|D,R) \label{5.3}
\eeq 
The potential $\Phi(u|D,R)$ of the Egoroff metric $H_i(u|D,R,r')$ is equal to 
\beq 
\Phi=\sum_{\alpha=1}^n\lambda(R_{\alpha}) \ d_{\alpha}\  
\psi(u,R_{\alpha}^{\sigma}), \label{5.3a}
\eeq 
where $d_{\alpha}$ are residues of $d\Omega_0$ at $R_{\alpha}$ (\ref{2.640}).
 
\end{th} 
{\it Proof.} The equality 
(\ref{5.3}) can be obtained by the consideration of the differential 
\beq
\lambda(Q)\Psi_i^0(u,Q)\Psi_j^0(u,\sigma(Q))d\Omega_0 \label{5.4} \eeq
that has poles at the points $P_i$ and $P_j$, only. The residues at these points 
are equal to $\beta_{ji}$ and $-\beta_{ij}$, respectively.

We would like to mention the other way to prove (\ref{5.3}), as well.
Let $\Psi_i^0$ and $\Psi_i^1$ be the Baker-Akhiezer functions 
defined by the conditions $1^{*}$-$4^{*}$. Then 
\beq 
\Psi_i^1(u,Q)=\lambda(Q)\Psi_i^0(u,Q) \label{5.5} 
\eeq 
Indeed, the function $\lambda \Psi_i^0$ satisfies all the defining conditions 
for the function $\Psi_i^1$. Therefore, (\ref{5.5}) is a corollary of the
uniqueness of the Baker-Akhiezer functions. From (\ref{5.5}) and (\ref{11})
it follows that the functions $\Psi_i^0$ satisfy the equations 
\bea 
\p_i \Psi^0_j & = & \beta_{ji} \Psi^0_i, \nonumber\\ 
\p_j \Psi^0_j & = & \lambda \Psi_j^1-\sum_{m\neq j} \beta_{mj} \Psi^0_m
\label{5.6} 
\eea 
The compatibility conditions of this system are equations (\ref{3}) and 
(\ref{4}) together with the symmetry condition (\ref{5.3}).  Note that 
(\ref{5.6}) as the auxiliary linear system for the full set of equations 
describing Egoroff metrics was proposed in \cite{dub}.

For the proof of (\ref{5.3a}) let us consider the differential
\beq
d\Omega_i^{(4)}=\lambda(Q) \psi(u,Q)\p_i \psi(u,\sigma(Q))d\Omega_0 \label{5.7}
\eeq
This differential has poles at the points $P_i$ and $R_{\alpha}$ with
the residues
\beq
{\rm res}_{P_i}d\Omega_i^{(4)}=-H_i^2,\ \
{\rm res}_{R_{\alpha}}d\Omega_i^{(3)}=d_{\alpha}\ \lambda(R_{\alpha}) \ 
\p_i \psi(u,R_{\alpha}^{\sigma}) \label{5.8}
\eeq
The sum of these residues is equal to zero. Therefore, (\ref{5.3a}) is proved.
\begin{cor} The formula
\beq
\Phi=\sum_{\beta=1}^ld_{\beta}\lambda(R_{\beta}) \Phi_{\beta}(u), \label{5.8a}
\eeq
where
\beq
\Phi_{\beta}=\sum_{\alpha=1}^l
r_{\alpha}(R_{\beta}^{\s})\ {\theta (A(R^{\s}_{\beta})+
\sum_{i=1}^n (u^i  V^{(i)})+Z_{\alpha}) \theta(Z_0)
\over \theta (A(R_{\beta}^{\s})+Z_{\alpha}) 
\theta (\sum_{i=1}^n (u^i V^{(i)})+Z_0) } 
\exp{\left(\sum_{i=1}^n 
u^i\int_{R_{\alpha}}^{\sigma(R_{\alpha})}d\Omega_i\right)}, 
\label{5.8b}
\eeq
defines the potentials of the algebraic-geometric Egoroff metrics.
\end{cor}

\section{\bf Solutions to the associativity equations} 

The equivalence of the classification problem for the rotation coefficients
of Egoroff metrics and the classification problem for the massive topological 
fields found in \cite{dub} does not provide the exact expression for the 
partition functions of these models. Now for each of the above-constructed
algebraic-geometrical Egoroff metrics we are going to define a function $F$ 
such that its third derivatives have the form
\beq
c_{klm}=\sum_{i=1}^n H_i^2\ {\p u^i\over \p x^k}{\p u^i\over \p x^l} 
{\p u^i\over \p x^m} \label{5.9}
\eeq
and the functions
\beq
c_{kl}^m= \sum_{i=1}^n {\p u^i\over \p x^k}{\p u^i\over \p x^l} 
{\p x^m\over \p u^i} \label{5.9a}
\eeq
define an associate algebra.
\begin{th} Let $\psi(u,Q|D,R)$ be the Baker-Akhiezer function
defined on the Riemann surface $\G$ of the function $\lambda(Q)$ and
corresponding to an admissible pair of the divisor $D,R$. Then the function 
$F(x)=F(u(x))$ defined by the formula 
\beq
F(u)={1\over 2} \left(\sum_{k,l=1}^n \eta_{kl} x^k(u) y^l(u)-
\sum_{\alpha}{d_{\alpha}\over 
\lambda(R_{\alpha})}\psi(u,R_{\alpha}^{\s})\right), \label{5.11} 
\eeq
where $\eta_{kl}$ is a constant matrix defined by (\ref{2.10}), 
$d_{\alpha}$ are equal to (\ref{2.640}) and 
\beq x^k(u)=\psi(u,Q_k),\ \ y^k={d\psi\over d\lambda}(u,Q_k), \label{5.12} 
\eeq
satisfies the equation 
\beq
{\p^3 F(x)\over \p x^k \p x^l \p x^m}= c_{klm} \label{5.10} 
\eeq
where $c_{klm}$ are defined by (\ref{5.9}). The functions $c_{kl}^m$ defined by 
(\ref{5.9a}) satisfy the associativity equations (\ref{15}).
\end{th} 
\noindent{\it Proof.} Let us consider the functions
\beq \phi_k={\p \psi\over \p x^k},\ \ \phi_{kl}={\p^2 \psi \over\p x^k\p x^l}
\label{5.13} \eeq
At the puncture $P_i$ they have the form 
\bea
\phi_k&=&{\p u^i\over \p x^k}\lambda e^{\lambda u^i} (h_i+O(\lambda^{-1})),\\
\phi_{kl}&=&{\p u^i\over \p x^k}{\p u^i\over \p x^k} \lambda^2e^{\lambda u^i} 
(h_i+O(\lambda^{-1})) \label{5.14} 
\eea
Therefore, 
\beq
c_{klm}=\sum_{i=1}^n {\rm res}_{P_i} d\Omega_{k;lm}, \label{5.15} \eeq
where 
\beq
d\Omega_{k;lm}=\phi_{k}(u,Q)\phi_{lm} (u,Q^{\sigma}) {d\Omega_0\over \lambda(Q)}
\label{5.15a} 
\eeq
By the definition of the flat coordinates $x^k$ we have 
\beq
\phi_k(u,Q_m)=\delta_{km},\ \ \phi_{kl}(u,Q_m)=0 \label{5.16} 
\eeq
Therefore, the differential $d\Omega_{k;lm}$ outside the punctures $P_i$ has 
the pole at $Q_k$, only. Hence, 
\bea
c_{klm} & = & -\ {\rm res}_{Q_k}d\Omega_{k;lm}=
- {\rm res}_{Q_k}\phi_{lm}(u,\sigma(Q_k)){d\Omega_0 \over \lambda(Q)}\\ 
& = & -\ {\p^2\over \p x^l \p x^m} \left({\rm res}_{Q_k}\psi(u,\sigma(Q_k))
\right){d\Omega_0 \over \lambda(Q)}. \label{5.17} 
\eea
At the puncture $Q_k$ the function $\psi(u,\sigma(Q))$ has the form
\beq 
\psi(u,\sigma(Q))=x^{\sigma(k)}-y^{\sigma(k)}\lambda +O(\lambda^2)\label{5.17a}
\eeq
and 
\beq
d\Omega_0={d\lambda\over \lambda}(\eta_{k}+\eta_k^1\lambda+O(\lambda^2))
\label{5.17b}
\eeq
Therefore,
\beq
{\rm res}_{Q_k}\psi(u,\sigma(Q_k)){d\Omega_0\over \lambda}=
\eta_k^1 x^{\sigma(k)} -\eta_{k} y^{\sigma(k)} \label{5.18} 
\eeq
From the definition of $F$ it follows that 
\beq
2{\p \over \p x^k}F=\eta_{k}y^{\sigma(k)}+\sum_{l}^n \eta_{l} x^l
{\p y^{\sigma(l)}\over \p x^k}-
{d_{\alpha}\over \lambda(R_{\alpha})}{\p \psi(u,R_{\alpha})\over
\p x^k}  \label{5.19} 
\eeq 
Consider the differential 
\beq
d\Omega_{k}^{(5)}={\p \psi(u,Q)\over \p x^k}\psi(u,Q^{\sigma})\ 
{d\Omega_0\over \lambda(Q)} \label{5.20} 
\eeq
That meromorphic differential has poles at the punctures $Q_l$ and at
the points $R_{\alpha}^{\sigma}$, only.  Its 
residues at these points are equal to 
\beq
{\rm res}_{Q_l}d\Omega_{k}^{(5)}=\eta_l x^{\sigma(l)}{\p y^l\over \p x^k}+
\delta_{l,\sigma(k)}\left(\eta_{k}^1 x^{\sigma(k)}-\eta_{k} y^{\sigma(k)}
\right)
\label{5.21}
\eeq
\beq
{\rm res}_{R_{\alpha}^{\sigma}}d\Omega_{k}^{(5)}=-{d_{\alpha}\over 
\lambda(R_{\alpha})} 
{\p \psi(u,R_{\alpha}^{\s})\over\p x^k} \label{5.21a}
\eeq
Therefore,
\beq
\sum_{l=1}^n \eta_lx^l {\p y^{\sigma(l)}\over \p x^k}
-\sum_{\alpha}{d_{\alpha}\over \lambda(R_{\alpha})} 
{\p \psi(u,R_{\alpha})\over\p x^k}=
\eta_k y^{\sigma(k)}- \eta_k^1 x^{\sigma(k)}\label{5.22}
\eeq
Finally,
\beq
{\p \over \p x^k}F=\eta_{k}y^{\sigma(k)}-{1\over 2}\eta_k^1 x^{\sigma(k)} 
\label{5.23} 
\eeq 
The last equality and equalities (\ref{5.17}), (\ref{5.18}) imply (\ref{5.10}).

\begin{lem} Let $\psi(u,Q|D,R)$ be the Baker-Akhiezer 
function defined on the Riemann surface $\G$ of the function $\lambda(Q)$ and
corresponding to an admissible pair of the divisors $D,R$. 
Then it satisfies the equations
\beq
{\p^2\over \p x^k \p x^l} \psi-\lambda \sum_{m=1}^n c_{kl}^m 
{\p\over \p x^m}\psi_m=0
\label{5.24}
\eeq
\end{lem}
{\it Proof.} Consider the function $\tilde \psi$ defined by the left hand side
of (\ref{5.24}). Outside of the punctures $P_j$ it has poles at the points
the divisor $D$, only, and is equal to zero at the points $Q_j$.
From the definition of $c_{kl}^m$ it follows that in the expansion of 
$\tilde \psi$ at the points $P_i$ the factor in front of the exponent has the 
form $O(\lambda^{-1})$. Therefore, from the uniqueness of the Baker-Akhiezer 
function it follows that $\tilde \psi=0$.

The associativity equations (\ref{15}) for the functions $c_{kl}^m$ are
the compatibility conditions for the system (\ref{5.24}). 
Theorem is proved.   

\noindent{\bf Remark}. The equations (\ref{5.24}) can be written in the
vector form
\beq
{\p \over \p x^k}\tilde \Psi_l=\lambda \sum_{m=1}^n c_{kl}^m\tilde \Psi_m, 
\label{5.25} \eeq where
\beq \tilde \Psi_k={\p \psi\over \p x^k}  \label{5.26}
\eeq 
The system (\ref{5.25}) with symmetric coefficients $c_{kl}^m=c_{lk}^m$
was introduced in \cite{dub} as an auxiliary linear system for the 
associativity equation (\ref{15}).

Now we are going to consider the special case of our construction when
the divisor $R$ coincides with the divisor ${\cal Q}$ of the punctures $Q_j$. 
As it was mentioned in the remark before the theorem 2.2 the assumption that 
$R$ does not intersect with $Q_j$ we take for the simplicity of the formulae, 
only. 

In the case when $R={\cal Q}$ of the admissible divisors $D$ are defined as
follows. The divisor $D=\g_1+\cdots+\g_{g+n-1}$ is called admissible if
there exists a meromorphic diiferential $d\Omega_0$ on $\G_0$ with 
poles of the order $2$ at the points $Q_{2m-n+1},\ldots, Q_{2m}$ and 
poles of the order $3$ at the double zeros of $E(P)$ and such that
\beq
d\Omega_0(\pi(\g_s))=0 \label{5.261}
\eeq
The differential $d\Omega_0$ considered as differential on $\G$ has the
form
\beq 
d\Omega_0={d\lambda\over \lambda^3(P)}(\eta_k+O(\lambda) \label{5.260} 
\eeq 
at the punctures $Q_k, \ k=1\ldots,n$.
In the special case under consideration, the flat coordinates are 
defined not by the evaluation the Baker-Akhiezer function $\psi$ at the 
punctures $Q^k$ (where it equals $1$ now) but by taking the next term in the 
expansions.  
\begin{th} Let $\psi(x,Q|D,{\cal Q})$ be the Baker-Akhiezer 
function defined by an admissible divisor $D$ on the Riemann surface of the 
function $\lambda(Q)$.  Then the function $F(x)=F(u(x))$ 
\beq F(u)={1\over 2}\sum_{k=1}^n\eta_{k}x^k(u) y^{\s(k)}(u), \label{5.28} 
\eeq 
where 
\beq
\eta_k={\rm res}_{Q_k}\lambda^2d\Omega_0 \label{5.280} 
\eeq and $x^k(u)$ and $y^k(u)$ are defined from the expansion 
\beq
\psi=1+x^k(u)\lambda+y^k(u)\lambda^2+O(\lambda^3) \label{5.29} 
\eeq
is a solution of the associativity equations (\ref{13})-(\ref{16}), i.e.  it 
satisfies the equation (\ref{5.10}), where $c_{klm}$ are given by (\ref{5.9});
the functions $c_{kl}^m$ defined by (\ref{5.9a}) satisfy (\ref{15}) and at 
last the relation 
\beq 
\sum_{m=1}^n c_{klm}(u)=\eta_{kl} \label{5.30} 
\eeq 
is fulfilled.  
\end{th} 
{\it Proof.} The proof that the functions $x^k$ are flat 
coordinates for the diagonal metric with the Lam\'e coefficients 
$H_i=\varepsilon_i h_i(u)$, where $h_i(u)$ is the leading term in the 
expansion of the corresponding Baker-Akhiezer function at the puncture
$P_i$ is just the same as in the general case. The proof of all the 
other statements of the theorem but the last one is also almost
identical to the their proof in Theorem 5.1
The last statement of the theorem (\ref{5.31}) follows from:  
\begin{lem} 
Let $\psi$ be the Baker-Akhiezer function corresponding to the data specified by 
the assumptions of Theorem 4.3. Then it satisfies the equation 
\beq \sum_{s=1}^n 
{\p\over \p u^k}\psi=\lambda \psi \label{5.31} 
\eeq 
\end{lem} 
The left and the right hand sides of (\ref{5.31}) are regular outside the 
punctures $P_k$ and have the same leading terms in their expansions at $P_k$. 
Therefore, they are identically equal to each other.

The evaluation of (\ref{5.31}) at $Q_m$ gives the equality
\beq
\sum_{k=1}^n {\p x^m\over \p u^s}=1 \label{5.32}
\eeq
Hence,
\beq
\sum_{m=1}^n c_{klm}(u)=\sum_{i=1}^n H_i^2\ 
{\p u^i\over \p x^k}{\p u^i\over \p x^l}=\eta_{kl} \label{5.33}
\eeq
Theorem is proved.

The exact theta functional formula for the partition function $F$ can be
obtained by the simple substitution of the corresponding expressions for
the Baker-Akhiezer function. The formulae for the functions $x^k(u)$ in 
(\ref{5.28}) are slightly different from (\ref{2.17a}), (\ref{2.17}).
Namely,
\beq 
x^k(u)={\p\over \p A(Q_k))}\sum_{\alpha=1}^l r_{m}(A(Q_k))\ x_m^k(u), 
\label{5.34} 
\eeq 
where
\beq
x_{m}^k(u) = {\theta (A(Q_k)+\sum_{i=1}^n (u^i V^{(i)})+Z_{m})
\theta(Z_0) \over \theta (A(Q_k)+Z_{m}) 
\theta (\sum_{i=1}^n (u^i V^{(i)})+Z_{0}) } 
\exp{\left(\sum_{i=1}^n u^i\int_{Q_m}^{Q_k}d\Omega_i\right)}
\label{5.36} 
\eeq
The formulae for $y^k(u)$ can be obtained by taking the derivatives of 
the formulae in the right hand sides of (\ref{5.34}) with respect to the 
variables $A(Q_k)$. 

\bigskip

\noindent{\it Example. Elliptic solutions}

\medskip

\noindent
Let us consider the simplest elliptic curvilinear coordinates and solutions 
to the associativity equations that correspond to $n=l=3,\ m=2$ 
in the example of Section 2.

Let us represent the elliptic curve $\G$ as a factor of the
complex plane of a variable $z$ with respect to the lattice
generated by periods $2\omega, 2\omega', \ {\rm Im}\ \omega'/\omega >0$. 
In that realization we identify the punctures $P_i$ with half periods $\o_i$, 
i.e. with the points: 
\beq P_1=\o_1=\omega,\ P_2=\o_2\omega',\ P_3=\o_3=-\omega-\omega'  \label{a1} 
\eeq 
The punctures $Q_j$ are the points 
\beq
Q_1=0, Q_2=z_0,\ \ Q_3=-z_0 \label{a2}
\eeq
In the case $g=1$ any sets $\g_1,\ldots,\g_l$ and $R_1,\ldots,R_l$ are the
admissible divisors. The corresponding differential
$d\Omega_0$ has the form
\beq
d\Omega_0=\eta_0{\s(z-\o)\s(z-\o')\s(z+\o+\o')
\over \s(z)\s(z+z_0)\s(z-z_0)}\prod_{s=1}^l {\s(z-\g_s)\s(z+\g_s)\over
\s(z-R_s)\s(z+R_s)}dz ,  
\label{a3}
\eeq
where $\s(z)=\s(z|\o,\o')$ is classical $\s$-Weierstrass function
The residues of this differential
\beq 
{\rm res}_{z=0}d\Omega_0=\eta_1,  \ {\rm res}_{z=\pm z_0}d\Omega_0=\eta_2
\label{a4}
\eeq
are the coefficients of the flat metric
\beq
ds^2=\eta_1 (dx^1)^2+\eta_2 (dx^2)(dx^3) \label{a41}
\eeq
The Baker-Akhiezer function corresponding to $\g_s$ and $R_s$  has the form:
 \beq
\psi(u,z)=\prod_{s=1}^l{\s(z-R_s)\over \s(z-\g_s)}
\left[\sum_{\alpha=1}^l r_{\alpha} 
{\s(z+U-R_{\alpha})\over \s(z-R_{\alpha})\s(U)} 
\exp(\Omega(u,z)-\Omega(u,R_{\alpha}))\right] , \label{a6}
\eeq
where 
\beq
U=u^1+u^2+u^3,\label{a61}
\eeq
\beq
\Omega(u,z)=u^1(\zeta(z-\o)+\eta)+u^2(\zeta (z-\o')+\eta')+
u^3(\zeta(z+\o+\o')-\eta-\eta'),
\label{a7}
\eeq
\beq
\zeta(z)={\s'(z)\over \s(z)},\ \ 
\eta=\zeta(\o), \ \eta'=\zeta(\o')\label{a70}
\eeq
and 
the constants $r_{\alpha}$ are equal to
\beq
r_{\alpha}={\prod_{s=1}^l\s(R_{\alpha}-\g_s)\over \prod_{s\neq \alpha}
\s(R_{\alpha}-R_s)} \label{a71}
\eeq
In the general case when $R_{\alpha}\neq Q_j$ the evaluation of $\psi$ at
$Q_j$ gives an expression of the flat coordinates through the curvilinear 
3-orthogonal coordinates  $u^1,u^2,u^3$:
\beq 
x^1=\psi(u,0),\ x^2=\psi(u,z_0),\ x^3=\psi(u,-z_0) \label{a9} 
\eeq 
The metric (\ref{a41}) in the coordinates $u^i$ is diagonal with the
coefficients
\bea
H_1(u)&=&\varepsilon_1\sum_{\alpha=1}^n r_{\alpha} 
{\s(U-R_{\alpha})\over \s(\o-R_{\alpha})\s(U)}e^{U\eta}\\
H_2(u)&=&\varepsilon_2\sum_{\alpha=1}^n r_{\alpha} 
{\s(\o'+U-R_{\alpha})\over \s(\o'-R_{\alpha})\s(U)}e^{U\eta'}\\
H_3(u)&=&\varepsilon_3\sum_{\alpha=1}^n r_{\alpha} {\s(\o+\o'+U-R_{\alpha})
\over 
\s(\o+\o'-R_{\alpha})\s(U)}e^{(-U\eta-U\eta')}. \label{a10} 
\eea
The constants $\varepsilon_i$ are equal to:
\bea
\varepsilon^2_1&=&\eta_0\ F(\o)\ {\s(\o-\o')\s(2\o+\o')
\over \s(\o)\s(\o+z_0)\s(\o-z_0)},\\
\varepsilon^2_2&=&\eta_0\ F(\o')\ {\s(\o'-\o)\s(2\o'+\o)
\over \s(\o')\s(\o'+z_0)\s(\o'-z_0)},\\
\varepsilon^2_3&=&-\eta_0\ F(\o+\o')\ {\s(2\o+\o')\s(2\o'+\o)
\over \s(\o+\o')\s(\o+\o'+z_0)\s(\o+\o'-z_0)},         \label{a11}
\eea
\beq
F(z)=\prod_{s=1}^l{\s(z-\g_s)\s(z+\g_s)\over
\s(z-R_s)\s(z+R_s)} \label{a110}
\eeq
The elliptic solutions for the associativity equations correspond to 
the Baker-Akhiezer functions given 
by the formula (\ref{a6}) with $l=3$ and $R_1=0, \ R_2=z_0, \ R_3=-z_0$.  
For the simplicity of the formulae let us consider the special case
corresponding to the basic elliptic Baker-Akhiezer function
\beq
\psi(u,z)={\s(z+s)\over \s(z)\s(s)}e^{\Omega(u,z)} \label{a12}
\eeq
The coefficients of the expansions of $\psi$ at the points 
$z=0,\ z_0,\ -z_0$ 
\bea
\psi&=&{1\over z}+x^1(u)+y^1(u)z+O(z^2),\\
\psi&=&x^2+y^2(u)(z-z_0)+O((z-z_0)^2),\\
\psi&=&x^3+y^3(u)(z+z_0)+O((z+z_0)^2) \label{a120}
\eea
define the solution 
\beq
F=x^1y^1-{1\over 2}(x^2y^3+x^3y^2)\label{a121}
\eeq
to the associativity equations. From (\ref{a12}) it follows that
\bea 
x^1&=&\zeta(U)-\wp(\o)u^1-\wp(\o')u^2-\wp(\o+\o')u^3,\\ 
x^2&=&{\s(z_0+U)\over \s(z_0)\s(U)}\exp\Omega(u,z_0),\\
x^3&=&{\s(U-z_0)\over \s(-z_0)\s(U)}\exp\Omega(u,-z_0) \label{a.13}
\eea
and
\bea 
y^1&=&{\s''(U)\over 2\s(U)}-\zeta(U)\sum_{i=1}^3(\wp(\o_i)u^i)+
{1\over 2}\ \left(\sum_{i=1}^3(\wp(\o_i)u^i)\right)^2,\\ 
y^2&=&
x^2(u)\left(\zeta(z_0+U)-\zeta(z_0)-\sum_{i=1}^3(\wp(z_0-\o_i)u^i\right),\\
y^3&=&
x^3(u)\left(\zeta(-z_0+U)+\zeta(z_0)-\sum_{i=1}^3(\wp(z_0-\o_i)u^i\right)
\label{a.131}
\eea
The function 
\beq
\tilde F= F-{1\over 2}(x^1)^2, \label{a14}
\eeq
has the same third derivatives as $F$. After substitution of the exact 
expression
for $x^i$ and $y^i$ into (\ref{a121}) we obtain the final formula for  
the simplest elliptic solution to the associativity equations
\beq
\tilde F=-{1\over 2}\wp(U)-{1\over 2}
\left(\wp(U)-\wp(z_0)\right)\left(\zeta(z_0-U)-\zeta(z_0-U)- 
\sum_{i=1}^3(\wp(z_0-\o_i)u^i)\right) \label{a15}
\eeq

\noindent{\bf Aknowledgements.} The author would like to thank
V. Zakharov and S. Manakov for their explanation of the ideas of their 
remarkable works that were an inspiration for this paper.


\begin{thebibliography}{**} 

\bibitem{dar}
G. Darboux, {\it Lecons sur le Systems Ortogonaux et les Coordones Curvilignes},
Paris (1910).

\bibitem{dn1}
B. Dubrovin, S. Novikov, {\it The Hamiltonian formalism of one dimensional
systems of hydrodynamic type and Bogolyubov-Whitham averaging method},
Sov. Math. Diklady 27. 1983,  665-654.

\bibitem{dn2}
B. Dubrovin, S. Novikov, {\it Hydrodynamics of of weakly deformed soliton
lattices: Differential geometry and Hamiltonian theory}, Russian Math Surveys
44, 1989, 35-124.

\bibitem{tzar}
S. Tsarev, {\it The geometry of Hamiltonian systems of hydrodynamic type.
Generalized Hodograph method}, Math in the USSR Izvestiya, 37, 1991, n 2,
397-419. 

\bibitem{dub}
B. Dubrovin, Nucl. Phys. B 379, 1992,  627.

\bibitem{kr_t}
I.Krichever, {\it Tau-function of the universal Whitham hierarchy and	
topological field theories}, 
Communications on Pure and Appl. Math., v. 47 (1994), 1-40.


\bibitem{zak}
V. Zakharov, {\it Description of the n-ortogonal curvilinear coordinate systems 
and hamiltonian integrable systems of hydrodynamic type. Part 1. Integration of 
the Lam\'e equations}, preprint

\bibitem{zm}
V. Zakharov, Manakov, private communication

\bibitem{wit}
E. Witten, {\it The structure of the topological phase of two-dimensional
gravity}, Nucl. Phys. B 340, 1990, 281-310.

\bibitem{vvd}
E. Verlinder, H. Verlinder, {\it A solution of two-dimensional topological
quantim gravity}, preprint IASSNS-HEP 90/40, PUPT-1176 ,1990.

\bibitem{kr1}
I.Krichever,{\it The algebraic-geometrical construction of Zakharov-Shabat 
equations and their periodic solutions},
Doklady Acad. Nauk USSR 227 (1976), n 2, 291-294.

\bibitem{kr2}
I.Krichever.{\it The integration of non-linear equations with the help of
algebraic-geometrical methods}
Funk.anal. i pril. 11 (1977), n 1, 15-31.

\bibitem{mi}
A. Its, V. Matveev, {\it On a class of Solutions of the Korteweg-de Vries 
equation}, Problem Mat. Fiz., 8, 1976, Izdat. Leningrad University,
Leningrad.

\bibitem{bab}
I. Krichever, O. Babelon, E. Billey, M. Talon,
{\it Spin generalization of the Calogero-Moser system and the Matrix
KP equation}, Amer. Math. Soc. Transl. (2), v. 170 (1995), 83-119.

\bibitem{nv}
S. Novikov, A. Veselov, {\it Finite-gap two-dimensional periodic 
Schr\"odinger operators: exact formulae and evolution equations}, Dokladi AN 
USSR, 279, 1984, n 1, 20-24. 

\bibitem{sit}
I. Krichever, {\it Algebraic-geometrical two-dimensional operators with 
self-consistent potentials},
Funk.anal. i pril., v.28 (1994), n 1, 26-40.

\bibitem{ndk}
B. Dubrovin, I.Krichever, S. Novikov {\it The Schr\"odinger equation in 
a periodic field and Riemann surfaces},
Doklady Acad. Nauk USSR 229 ,1976, n 1, 15-18.


\end{thebibliography}
\end{document}